\documentclass[traditabstract,epsf]{aa}

\def\m2s2{\hbox{\,m$^{2}$\,s$^{-2}$}} 

\usepackage{graphicx}
\usepackage{txfonts}
\usepackage{verbatim}
\begin{document}

 \title{Fast-evolving weather for the coolest of our two new substellar neighbours}
 
\author{
          M.~Gillon\inst{1},
          A.~H.~M.~J. Triaud\inst{2, 3},
          E.~Jehin\inst{1}, 
          L.~Delrez\inst{1},
          C.~Opitom\inst{1},
          P.~Magain\inst{1},
          M.~Lendl\inst{4},
          D.~Queloz\inst{4}
           }

\offprints{michael.gillon@ulg.ac.be}
\institute{ 
       	    $^1$ Institut d'Astrophysique et G\'eophysique, Universit\'{e} de Li\`{e}ge, all\'{e}e du 6 Ao\^{u}t 17, B-4000 Li\`{e}ge, Belgium\\
	    $^2$ Department of Physics, and Kavli Institute for Astrophysics and Space Research, Massachusetts Institute of Technology, Cambridge, MA 02139, USA\\
	    $^3$ Fellow of the Swiss National Science Foundation\\
	    $^4$ Observatoire de Gen\`eve, Universit\'e de Gen\`eve, 51 Chemin des Maillettes, 1290 Sauverny, Switzerland}

\date{Received date / accepted date}
\authorrunning{M. Gillon et al.}
\titlerunning{Fast-evolving weather for Luhman\,16B}

  \abstract
  {We present the results of an intense photometric monitoring in the near-infrared ($\sim$0.9 $\mu$m) with 
  the TRAPPIST robotic telescope of the newly discovered  binary brown dwarf WISE\,J104915.57-531906.1,
    the third closest system to the Sun at a distance of only 2 pc.
     Our twelve nights of photometric time-series reveal a quasi-periodic ($P = 4.87 \pm 
   0.01$h) variability  with a 
  maximal peak-peak amplitude of $\sim$11\% and strong night-to-night evolution. We  attribute this variability 
  to the rotational modulation of fast-evolving weather patterns in the atmosphere of  the coolest 
  component ($\sim$T1-type) of the binary, in agreement with the cloud fragmentation mechanism proposed to drive the spectroscopic morphologies
  of  brown dwarfs at the L/T transition. No periodic signal is detected for the hottest component ($\sim$L8-type).
   For both brown dwarfs, 
  our data allow us to firmly discard any  unique transit during our observations for planets $\ge 2  R_\oplus$. 
  For orbital periods  smaller than $\sim$9.5h,  transiting planets are excluded down to an Earth-size.  }


   \maketitle

\section{Introduction}

Both inaugurated in 1995 (Rebolo et al. 1995; Mayor \& Queloz 1995), the observational studies of 
brown dwarfs (BD) and exoplanets are two of the most active fields of modern astronomy. The 
atmospheric study of these substellar objects has seen tremendous advances , thanks to
the sophistication of models and the constant improvement of instruments (see, e.g., Seager \& Deming 
2010 and Showman \& Kaspi 2012 for reviews). With effective temperatures ranging from $\sim$300-2000 K, 
L and T-types field BDs  amenable for detailed direct studies represent critical precursors to  the 
atmospheric characterization of giant exoplanets. The data 
gathered so far outline the important role of atmospheric condensates on the spectral 
morphologies of these objects (Kirckpatrick 2005).  This is especially true at the L-T transition ($\sim$L7-T4 
spectral types) which is characterized by an increase of the $J$-band luminosity with decreasing temperature 
(Vrba et al. 2004). This has been explained by the gradual depletion of silicates in the cooler atmospheres resulting in 
increasingly patchy cloud covers and thus increasingly small condensate opacity. Still, the details of this 
transition remain poorly understood (e.g. Saumon \& Marley 2008).  Ackerman \& Marley (2001) have proposed
 to explain the condensate depletion by the fragmentation of the clouds  driven by convection in the troposphere. 
 This scenario predicts relatively large 
($\sim$1-20\%) photometric variability around $1\mu$m on rotational timescales, driven by the 
formation, evolution and complex dynamics of cloud holes in the upper atmosphere. 

Brown dwarfs being very rapid rotators (typically a few hours, Herbst et al. 2007), this variability is observable 
within a few nights of photometric monitoring. Because of the extreme faintness 
in the optical at the L/T transition, only a handful of near-infrared (NIR) monitoring using medium-sized 
ground-based telescopes or space-based facilities could reach the photometric precisions required to
detect these predicted semi-periodic variabilities (Clarke et al. 2008; Artigau et al. 2009; 
Radigan et al. 2012; Khandrika et al. 2013; Apai et al. 2013). Some of these results agree nicely
with  the cloud fragmentation hypothesis, but there is still no clear evidence that BDs at the L/T transition are
more variable than the other BDs (Khandrika et al. 2013). More high-precision photometric monitoring 
of L/T transition BDs are thus highly desirable. High-precision time-series can also inform us on the 
spatial and temporal distribution of cloud structures,  vertical thermal profiles, degrees of differential rotation, 
and on the age of brown dwarfs since their rotation period decreases 
monotonically with time.

A unique opportunity for this domain came up recently with the detection by Luhman (2013, hereafter L13) 
of a nearby binary BD at only $2.02\pm0.15$ pc. This system, WISE\,J104915.57-531906.1 (hereafter
Luhman\,16, following Mamajek 2013), is the third closest system to Earth, making it an exquisite target for high
signal-to-noise ratio follow-up ($J=10.7$, $K=8.8$). With spectral types L8$\pm$1 for Luhman\,16A and 
T1$\pm$2 for Luhman\,16B (Kniazev et al. 2013, hereafter K13; see also Burgasser et al. 2013), both of 
its components span the L/T transition. This system is an invaluable target for high-precision time-series photometry, 
an interest further reinforced
by the classification of the pair as a possible variable in WISE All-Sky Source Catalog (L13). This
motivated us to perform an intensive photometric monitoring of the system in the NIR using the 60cm robotic 
telescope TRAPPIST. We observed Luhman\,16 for twelve nights, 
reaching a precision that allowed us to detect  a clear quasi-periodic variability,  
attributed to  the T-dwarf Luhman\,16B, and combined with a fast evolution of the observed patterns. 

The next section  presents our TRAPPIST data and their reduction. Our analysis of the resulting photometric 
time-series is described in Sec.~3.  Finally, we discuss briefly our results in Sec.~4. 

\section{Data description}

We monitored WISE\,1049-5319 for twelve nights between 2013 March 14 and 26 
with the robotic 60cm telescope TRAPPIST ({\it TRA}nsiting {\it P}lanets and {\it P}lanetes{\it I}mals 
{\it S}mall {\it T}elescope; Gillon et al. 2011, Jehin et al. 2011) located at ESO La Silla Observatory 
in the Atacama Desert, Chile. TRAPPIST is equipped with a thermoelectrically-cooled 2K\,$\times$\,2K 
CCD having a pixel scale of 0.65'' that translates into a 22'\,$\times$\,22'  field of view. The observations 
were obtained with an exposure time of 115s, with the telescope focused and through a special `$I+z$' filter 
having a transmittance $>$90\% from 750 nm to beyond 1100 nm\footnote{http://www.astrodon.com/products/filters/near-infrared/}.  
Considering the transmission curve of this $I+z$ filter,  the spectral response curve of the CCD detector, 
and the spectral type of the target, we derive an effective wavelength $\sim$910nm for the observations. 
The positions of the stars on the chip were maintained to within a few pixels over the course of 
each run, thanks to a `software guiding' system deriving regularly  an astrometric solution for the 
most recently acquired image and sending pointing corrections to the mount if needed.  

After a standard pre-reduction (bias, dark,  flatfield correction), the stellar fluxes for each run
 were extracted from the images using the {\tt IRAF/DAOPHOT}\footnote{{\tt IRAF} is 
 distributed by the National Optical Astronomy Observatory, which is 
 operated by the Association of Universities for Research in Astronomy, Inc., under cooperative agreement
  with the National Science Foundation.} aperture photometry software (Stetson, 
 1987). The same photometric aperture of 8 pixels ($\sim$5.1'') was used for all nights.  After a careful
  selection of stable reference stars of similar brightness, differential photometry was then obtained. 
  Finally, the  light curves were normalized. The twelve resulting light curves are shown in Fig.~1. To assess the 
  night-to-night variability of the target, we also extracted a global differential light curve that is shown in Fig.~2.
 
 The typical full-width at half maximum of TRAPPIST point-spread function (PSF) is $\sim$3 pixels = $\sim$2''. 
At this resolution, the two components of the 1.5'' binary are only partially resolved, and our photometry 
extracted with an aperture of $\sim$5.1'' radius shows thus the evolution of the sum of the fluxes of both components.

\begin{figure}[h!]
\label{fig:1}
\centering                     
\includegraphics[width=8.5cm]{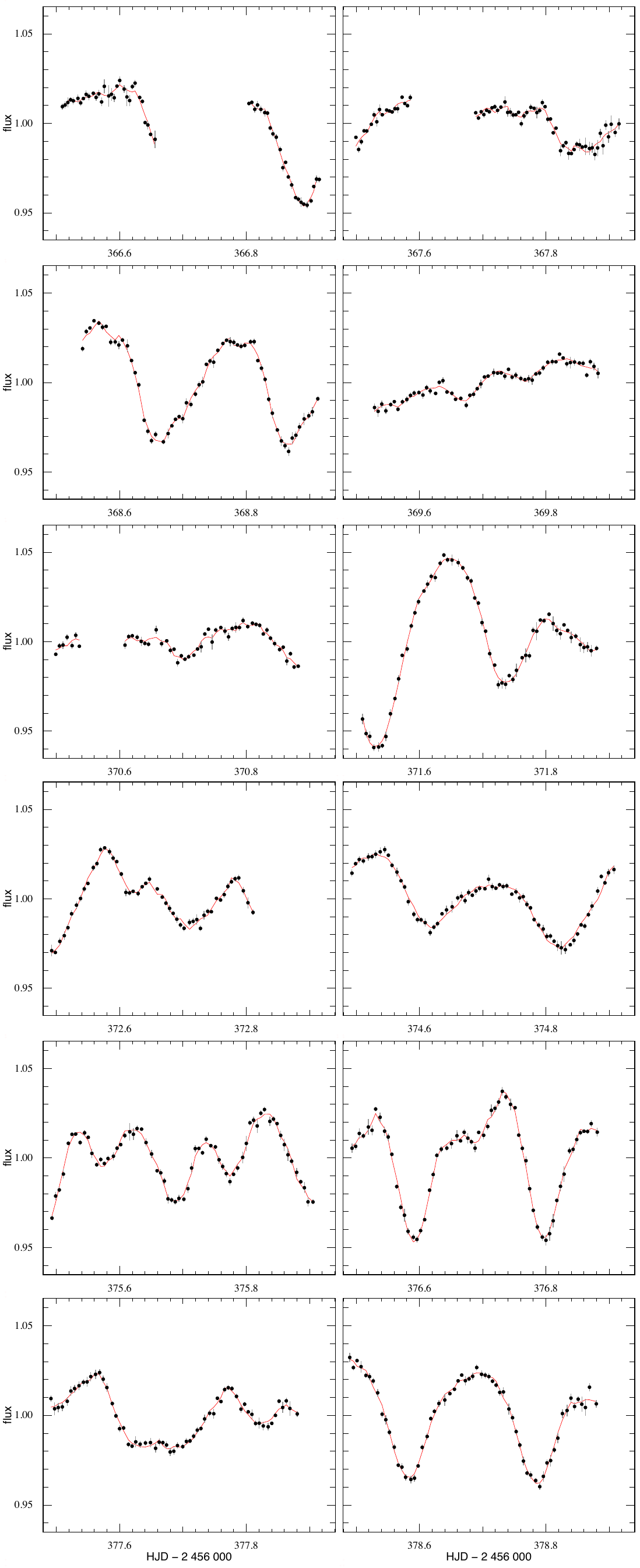}
\caption{Normalized TRAPPIST light curves for Luhman\,16 binned per 10 min intervals.
For each light curve, the best-fit global model (see Sec.~3) is over-imposed in red. Gaps for night \#1
and \#5 correspond to cloudy conditions. The gap for night \#2 corresponds to the observation of another
target. }
\end{figure}

\begin{figure*}
\label{fig:2}
\centering                     
\includegraphics[width=18cm]{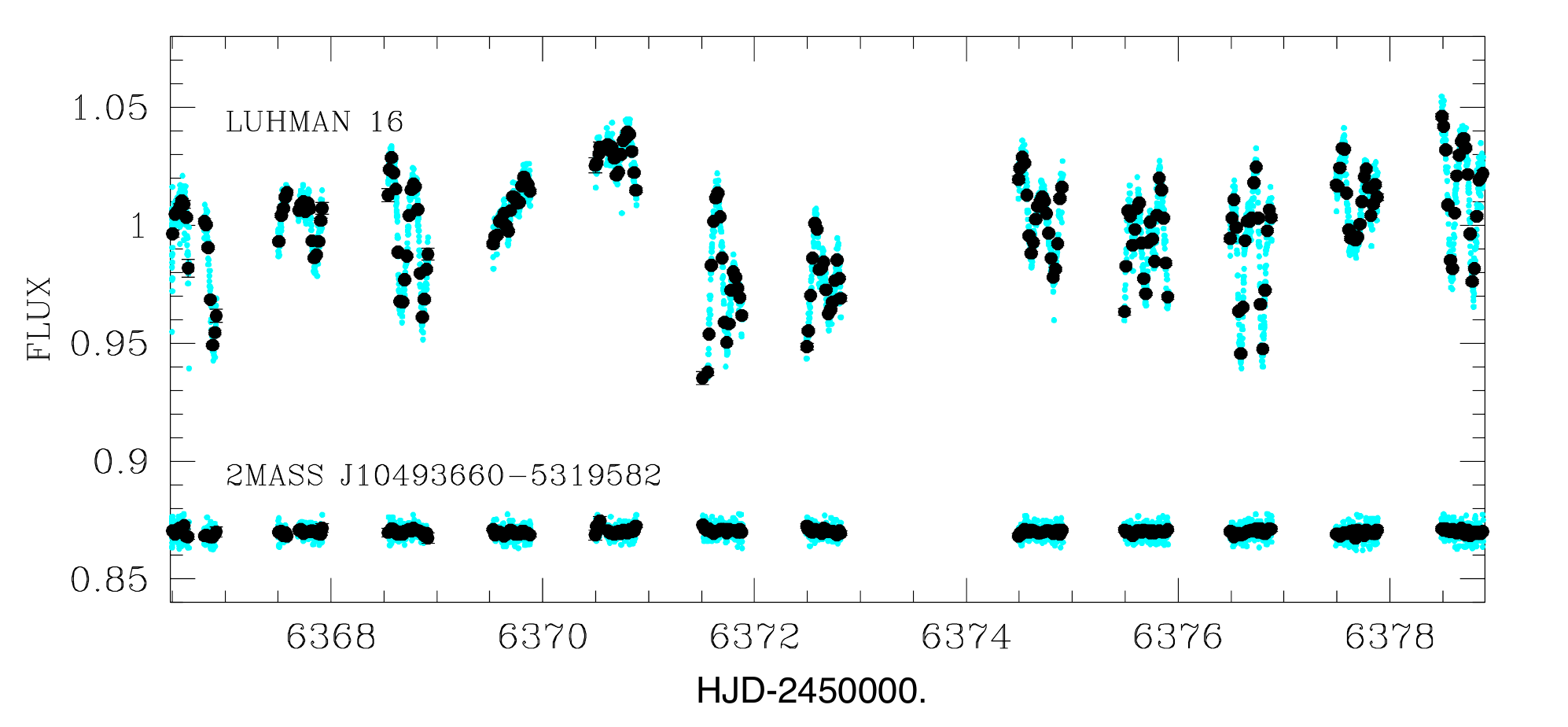}
\caption{Globally normalized TRAPPIST differential photometry for Luhman\,16 ($top$) and
for one of the comparison stars used in the reduction ($bottom$, shifted along the $y$-axis
for the sake of clarity),  unbinned ($cyan$) and
binned per interval of 30 min ($black$). The standard deviation of the binned light curves
are  2.2\% and  0.1\% for Luhman\,16 and the comparison star, respectively.}
\end{figure*}



\begin{figure}
\label{fig:3}
\centering                     
\includegraphics[width=8cm]{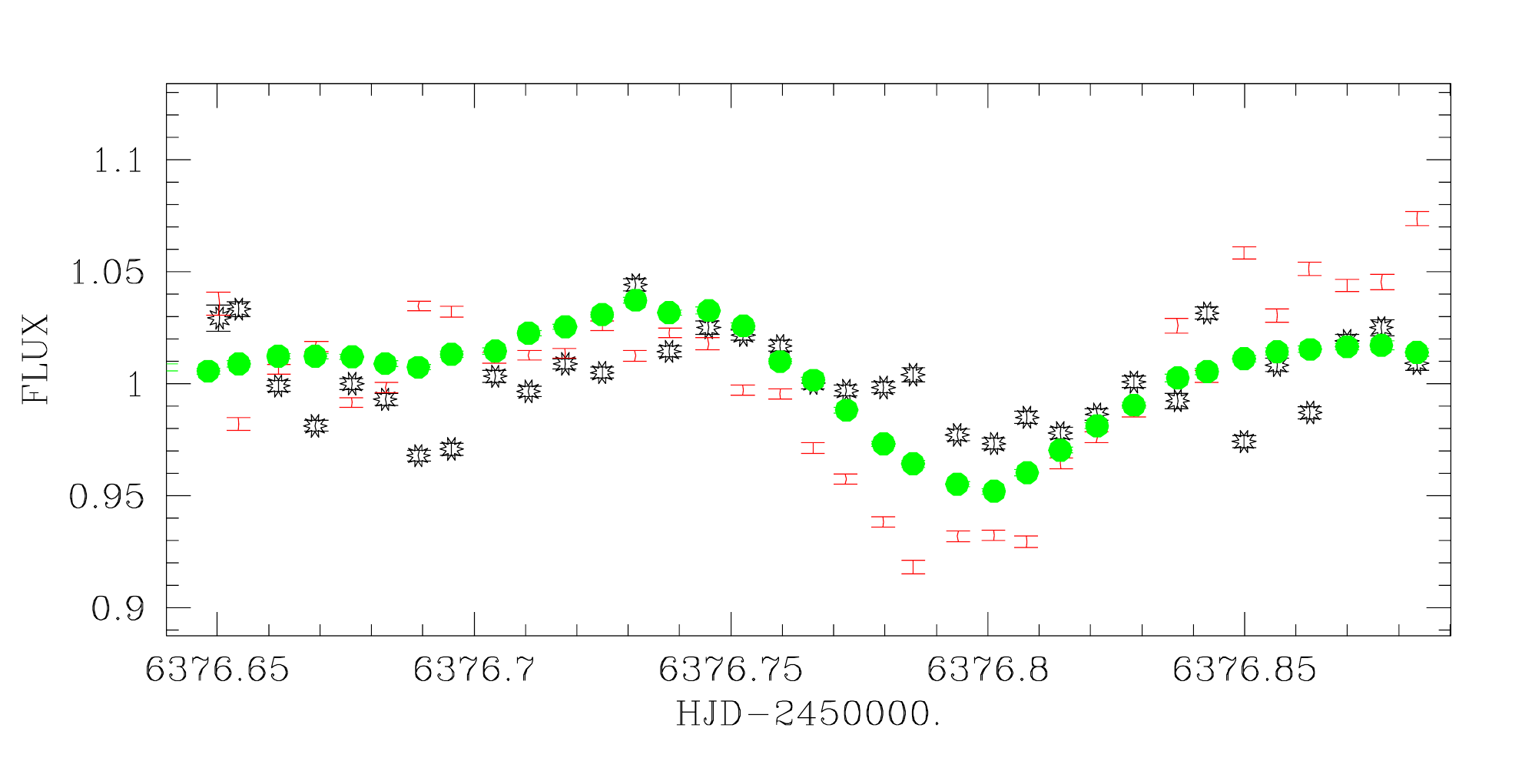}
\caption{Differential photometry for the second part of night \#10, binned per 10 min intervals, obtained with 
an aperture encompassing both components of the binary ({\it filled green circles}), 
and with apertures encompassing only the PSF center of Luhman\,16A ({\it black empty symbols}) 
and Luhman\,16B ({\it red vertical bars}). The amplitude of the variations is amplified on 
Luhman\,16B, indicating that they originate from this T-dwarf.}
\end{figure} 

\section{Data analysis}

While being relatively stable on longer timescales (Fig.~2), Luhman\,16 shows  a clear variability 
on a nightly timescale (Fig.~1). Furthermore, the observed patterns  evolve strongly from one night to the other.
A Lomb-Scargle periodogram (Lomb 1976; Scargle 1982) applied to our photometry shows a strong
power excess around $\sim$0.2d, matching well the typical separation between similar features in the 
light curves.
   
We  performed a global analysis of our twelve light curves, adapting for that purpose the Markov Chain Monte Carlo (MCMC)
code described in Gillon et al. (2012, hereafter G12). In a first step, we assumed that the observed variability 
was originating from only one of the two BDs. Our basic model
 for the rotational modulation was based on a division of the brown dwarf into 10 longitudinal slices. For each of them, 
 the surface flux was assumed to be constant during each night, the measured flux being modeled by a semi-sinusoidal
  function falling to zero when the slice disappears from view. To this rotational model, we added a baseline model 
  accounting for (i) the flip of the equatorial mount of the telescope at the meridian, putting the stellar images on 
  different pixels and thus possibly creating small offsets in the differential photometry, (ii) a 2$^{nd}$-order airmass 
  polynomial aiming to model the differential extinction curvature due to the much redder color of the target compared 
  to the comparison stars, and (iii) a  $4^{th}$-order time polynomial representing the low-frequency variability of the system, 
  including the evolution of the patterns from one rotation to the other. In this global model, the only two 
  perturbed parameters in the MCMC were the rotation period and an arbitrary phase, the solution for 
  the remaining parameters being obtained by linear regression at each step of the Markov chains (see G12 for details). 
Two MCMC chains of 100,000 steps were performed to probe efficiently the posterior distribution of the rotational period. 

Our MCMC analysis gives $P_{rot} = 4.87 \pm 0.01$h, producing an excellent fit between the model and the data (see Fig.~1). 
While a periodogram of the residuals reveals no additional period, we performed an additional analysis by adding a second 
rotational model to the MCMC. This analysis also failed to identify a second period. The Bayesian Information Criterion 
(Schwarz 1978) significantly increased by +360, indicating that a model including a single rotation period is a more likely 
representation of the data. We thus conclude that only one of the two BDs dominates the photometric variability since a same rotational 
period for both BDs is unlikely.

A new photometric reduction of a fraction of our images using the 
{\tt IRAF/DAOPHOT ALLSTAR} PSF-fitting software (Stetson 1987) revealed that
the B component is   -$0.1\pm0.1$ mag brighter than the A component

We then attempted to determine the origin of the signal. We extracted the fluxes of both components by aperture 
photometry, fixing the aperture centers on the known positions of the two BDs. The aperture sizes were chosen to 
encompass only one PSF center. Using only data showing both a large signal and having the smallest and most stable 
PSF widths, we find that a larger amplitude is visible on the T-dwarf, while no significant signal is  obtained for the L-dwarf. 
This is illustrated for a fraction of night \#10 in Fig.~3. From these results, we conclude that the detected quasi-periodic 
variability originates from the T-dwarf Luhman\,16B.

\section{Discussion}

From their new spectroscopy and from the relationship of Stephens et al. (2009),
K13 attribute to Luhman\,16A and B components effective temperatures of 1350$\pm$120K and  1220$\pm$110K, respectively.
Such very low temperatures make atmospheric condensates the most likely source of the observed rotational variability
(see discussion by Radigan et al. 2012 for the T1.5 BD 2M2139). The most surprising feature of the 
patterns reported here is their fast evolution from night to night. To our knowledge, this is the first time
that such rapid evolution is reported for the cloud coverage of an L/T transition BD, making Luhman\,16B
an actual `Rosetta Stone' for the study of BDs atmospheres. Future multi-band and high-cadence spectroscopic
 time-series will be able to observe while clouds condensate and are then dispersed, giving unprecedented access 
 into the physico-chemical processes acting in such atmospheres.

Considering our measured $\Delta$mag = $0.1\pm 0.1$ between both BDs, 
the variability amplitudes visible in Fig.~1 are thus diluted by a factor $\sim$2  by the blend of both PSFs in the
TRAPPIST images. The actual maximal peak-to-peak variability amplitude of Luhman\,16B  should thus be 
$>$20\%. Similar  amplitudes were observed around 1$\mu$m for at least another early T-dwarf, the T1.5-type BD 
2M2139 (Radigan et al. 2012). For the L-dwarf  Luhman\,16A, we do not detect any variability, 
but our  sensitivity is limited by the  partial resolution of both components in our images.
High-cadence photometry with  a better spatial resolution will be need to thoroughly assess its photometric variability.

Binary stars have been used from the dawn of modern astronomy as a means to control some variables, while leaving 
others free. In our case both BDs probably formed together out of the same fragment; they should have their 
age and composition on a par. What differentiates them is their mass. This affects their respective effective temperatures 
and densities. Although only 100 K apart, that one shows clouds breaking while the other does not is an indication of 
how sharp the boundary from a stable to an unstable atmosphere is, and of how closely tuned it must be to effective 
temperature and scale height.

Field BDs are very interesting targets for exoplanets transit searches (Blake et al. 2008; Bolmont et al. 2011;  
Belu et al. 2013), as their small sizes make possible the detection of terrestrial planets with photometric precisions 
at the mmag level similar to the ones reported here for Luhman\,16. A careful visual inspection of the 
residuals light curves and  a search for periodic box-like patterns with the BLS algorithm (Kovacs et al. 2002) 
failed to detect any transit-like signal. Despite the blend of both BDs, our sensitivity is good enough to partially probe the 
terrestrial regime. Fig.~4 (top panel) shows the residuals of our global modeling compared to the expected transit 
depths for  a 2 $R_\oplus$ radius planet,  for both BDs, assuming for each of them a 0.1$R_\odot$ radius. The corresponding
transit is firmly discarded by our data. For orbital periods smaller than the mean duration of our runs ($\sim$9.5h), our 
detection threshold goes down to Earth-size planets, as can be see in Fig.~4 (bottom panel). This shows that 
intensive campaigns like the one described here targeting the most nearby field BDs with small to medium-sized 
ground-based telescopes or, better, with an  infrared space facility  like {\it Spitzer} (Triaud et al. in prep.) could
 efficiently assess the frequency of close-in terrestrial planets around BDs, possibly detecting Earth-sized planets 
 amenable for atmospheric characterization with, e.g., JWST (Belu et al. 2013).

\begin{figure}
\label{fig:4}
\centering                     
\includegraphics[width=8cm]{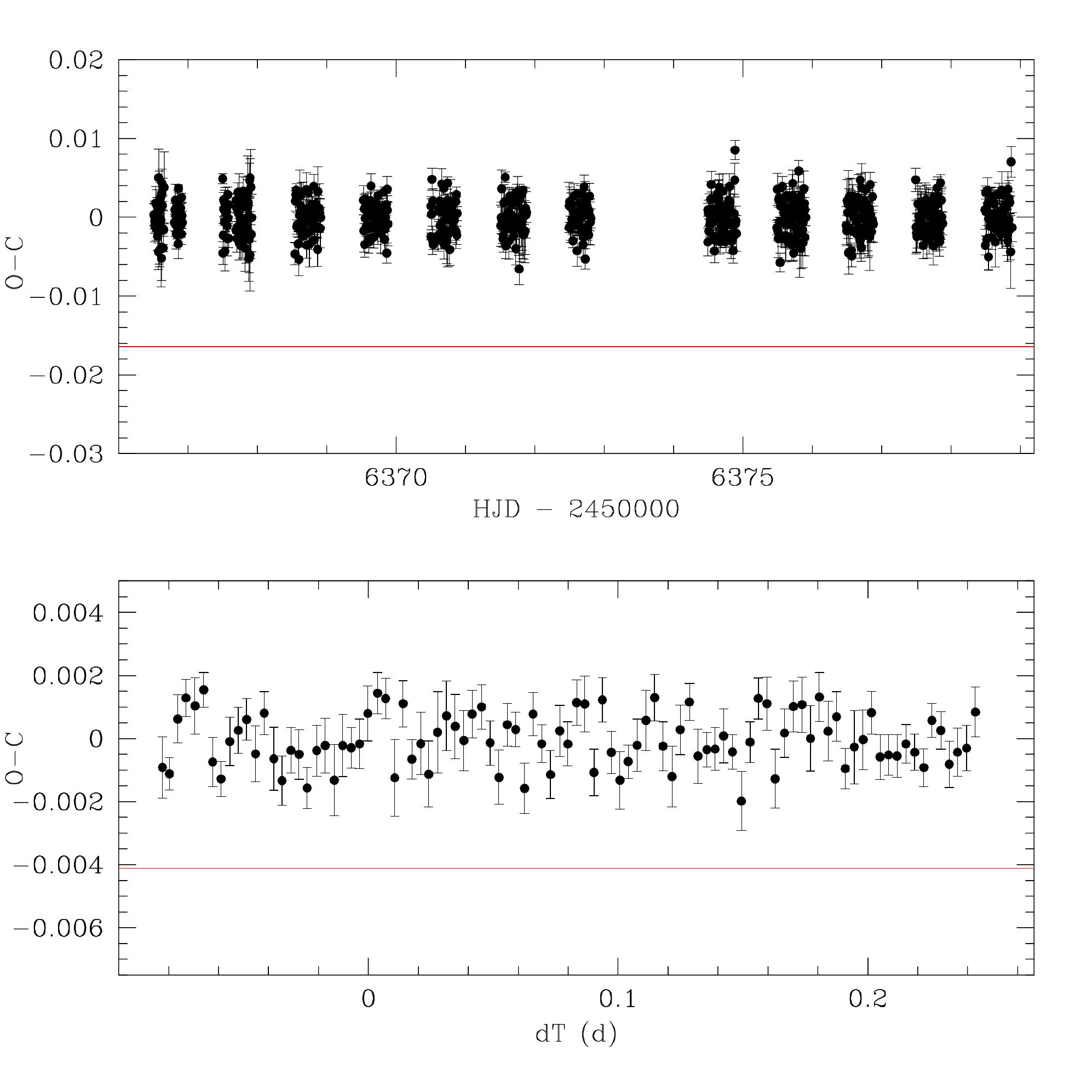}
\caption{Residuals of our global modeling, unfolded and binned per 10 min intervals ($top$), 
folded with $P=8$h and binned per 5 min intervals ($bottom$).  For the unfolded residuals
is shown as a red horizontal line the expected transit depth for a $2R_\oplus$ planet orbiting 
one of the two BD, assuming a 0.1$R_\odot$ radius and the same brightness for both BDs. 
The same is done for the folded residuals, but assuming a  $1R_\oplus$ planet.}
\end{figure}

\begin{acknowledgements}
The authors thank Mark S. Marley for his valuable suggestion. TRAPPIST is a project funded by the Belgian Fund for Scientific 
Research (Fonds National de la Recherche Scientifique, F.R.S-FNRS) under grant FRFC 2.5.594.09.F, with the participation of 
the Swiss National Science Foundation (SNF).  M. Gillon and E. Jehin are FNRS Research Associates. A.H.~M.~J. Triaud is Swiss 
National  Science Foundation fellow under the grant PBGEP2-145594. C. Opitom and L. Delrez thank the Belgian FNRS for funding 
their PhD theses. 
\end{acknowledgements}


\begin{thebibliography}{}

\bibitem{} Ackerman, A.~S., \& Marley, M. S. 2001, ApJ, 556, 872

\bibitem{} Artigau, E., Bouchard, S., Doyon, R., Lafreni\`ere, D. 2009, ApJ, 701, 1534
   
\bibitem{} Apai, D., Radigan, J., Buenzli, E., et al., 2013, ApJ (in press), arXiv:1303.4151

\bibitem{} Belu, A.~R., Selsis, F., Raymond, S.~N., et al. 2013, ApJ (in press), arXiv:1301.4453

\bibitem{} Bolmont, E., Raymond, S.~N., Leconte, J. 2011, A\&A, 535, 94

\bibitem{} Blake, C.~H., Bloom, J.~S., Latham, D.~W., et al. 2008, PASP, 120, 860


\bibitem{} Burgasser, A.~J., Sheppard, S.~S, \& Luhman, K.~L. 2013, ApJ (submitted), arXiv:1303.7283

\bibitem{} Clarke, F.~J., Hodgkin, S.~T., Oppenheimer, B.~R., et al. 2008, MNRAS, 386, 2009

\bibitem{} Gillon, M., Jehin, E., Magain, P., et al. 2011, {\it Detection and Dynamics of Transiting Exoplanets}, Proceedings of OHP Colloquium (23-27 August 2010), eds. F. Bouchy, R.~F. Diaz \& C. Moutou, Platypus Press, 06002

\bibitem{} Gillon M., Triaud A.~H.~M.~J., Fortney J.~J., et al. 2012, A\&A, 542, 4

\bibitem{} Herbst, W., Eisl\"offel, J., Mundt, R., \& Scholz, A. 2007, Protostars and Planets V, B. Reipurth, D. Jewitt, and K. Keil (eds.), University of Arizona Press, Tucson, 951, 297

\bibitem{} Khandrika, H., Burgasser, A.~J., Melis, C., et al. 2013, AJ, 145, 71

\bibitem{} Kovacs, G., Zucker, S., \& Mazeh, T. 2002, A\&A, 391, 369

\bibitem{} Jehin, E., Gillon, M., Queloz, D., et al. 2011, The Messenger, 145, 2

 \bibitem{} Kirkpatrick, J.~D. 2005, ARA\&A, 43, 195
 
 \bibitem{} Kniazev, A.~Y., Vaisanen, P., Muzic, K., et al., 2013, ApJL (submitted), arXiv1303.7171
 
 \bibitem{} Lomb, N.~R., 1976, Ap\&SS, 39, 447

\bibitem{} Luhman, K.~L. 2013, ApJL (in press), arXiv:1303.2401

\bibitem{} Mamajek, E.~E. 2013, arXiv1303.5345

\bibitem{} Mayor, M., \& Queloz D.  1995, Nature, 378, 355
  
\bibitem{} Rebolo, R., Zapatero Osorio, M.~R., \& Mart\'in, E.~L. 1995, Nature, 377, 129

\bibitem{} Radigan, J., Jayawardhana, R., Lafreni\`ere, D., et al. 2012, ApJ, 750, 105

\bibitem{} Saumon, D., \& Marley, M.~S. 2008, ApJ, 689, 1327

\bibitem{} Scargle, J.~D., 1982, ApJ, 263, 835

\bibitem{} Schwarz, G.~E. 1978, Annals of Statistics, 6, 461  

\bibitem{} Seager, S., \& Deming, D. 2010, Exoplanets Atmospheres, ARAA, 48, 631
    
\bibitem{} Showman, A. P., \& Kaspi, Y. 2012, ApJ (submitted), arXiv:1210.7573

\bibitem{} Stephens, D.~C., Leggett, S.~K., Cushing, M.~C., et al. 2009, ApJ, 702, 154
    
\bibitem{}  Stetson, P.~B.  1987, PASP, 99, 111

\bibitem{} Vrba, F. J., Henden, A.~A., Luginbuhl, C.~B., et al. 2004, AJ, 127, 2948

\end{thebibliography}
\end{document}